\documentclass[12pt]{article}

\usepackage{cite}

\usepackage[nointlimits,reqno]{amsmath}
\usepackage{amssymb}
\numberwithin{equation}{section} 

\usepackage{graphicx}
\usepackage{adjustbox}
\usepackage{paralist, tabularx}
\usepackage[toc,page]{appendix}

\usepackage{subcaption}
\usepackage{float}
\usepackage{caption}
\usepackage{luty}

\newcommand{\met}{\Sla{E}_{\rm T}}
\newcommand{\pt}{\ensuremath{p_{\rm T}}}

\newcommand{\ifb}{\mbox{ fb}^{-1}}

\begin{document}
\begin{titlepage}

\title{Probing Exotic Triple Higgs Couplings at the LHC}

\author{Christina Gao}

\address{Theoretical Physics Department, Fermilab, Batavia, IL 60510}

\author{Nicol\'as A. Neill}

\address{Instituto de Alta Investigaci\'{o}n, Universidad de Tarapac\'{a}, Casilla 7D, Arica, Chile}

\begin{abstract}
In extended Higgs sectors that exhibit alignment without decoupling, the additional scalars are allowed to have large couplings to the Standard Model Higgs.
We show that current nonresonant di-Higgs searches can be straightforwardly adapted to look for additional Higgses in these scenarios, where pair production of non-SM Higgses can be enhanced.
For concreteness, we study pair production of exotic Higgses in the context of an almost inert two Higgs doublet model, where alignment is explained through an approximate $\mathbb{Z}_2$ symmetry under which the additional scalars are odd. In this context, the smallness of the $\mathbb Z_2$ violating parameter suppresses single production of exotic Higgses, but it does not prevent a sizeable trilinear coupling $hHH$ between the SM Higgs ($h$) and the additional states ($H$).
We study the process $pp\rightarrow h^* \rightarrow HH$ in the final states $b\bar b b \bar b$, $b\bar b\gamma\gamma$, and multi-leptons. We find that at the HL-LHC these modes could be sensitive to masses of the additional neutral scalars in the range $130\mbox{ GeV} \lesssim m_H \lesssim 290\mbox{ GeV}$.
\end{abstract}
\end{titlepage}


\newcommand{\norm}[1]{|\!\gap|#1|\!\gap|}
\newcommand{\aavg}[1]{\avg{\!\avg{#1}\!}}

\section{Introduction}
The properties of the observed 125 GeV Higgs boson are in close agreement with the Standard Model (SM) predictions.
Therefore, any extension of the SM Higgs sector needs to reproduce the SM-like Higgs couplings to the weak gauge bosons and fermions.
In multi-Higgs sectors, one possible realization of this agreement is near the so-called ``decoupling limit'', where the physical masses of the additional Higgses are sent to infinity while keeping the remaining dimensionless self-couplings fixed \cite{Haber:1989xc,Gunion:2002zf}. 
Another possibility, the so-called ``alignment limit'', is that the vacuum expectation value (VEV) of one of the CP-even states is closely aligned with the direction of the VEVs in the multi-Higgs space \cite{Gunion:2002zf,Craig:2013hca,Carena:2013ooa,Delgado:2013zfa}.
In both limits the properties of one of the Higgses approach those of the SM Higgs.
The decoupling limit implies the alignment limit, but alignment can also be realized without requiring the additional scalars to be much heavier than the electroweak scale. 
In this kind of scenario we focus on in this work.

The agreement of the properties of the observed 125 GeV Higgs with the SM predictions can be easily satisfied in a setup where the observed Higgs boson comes mainly from a SM doublet $H_1$ plus a small deviation from it:
\[\label{ep}
H_{SM}\equiv H_1+\epsilon H_2.
\]
The second doublet $H_2$ introduces four new degrees of freedom, which can be represented by the following physical states: a charged Higgs $H^{\pm}$, a neutral scalar $\phi$, and a neutral pseudoscalar $A$.
In this scheme, they all have $\epsilon$-suppressed interactions with the SM particles, therefore they are difficult to look for experimentally.

A SM-like Higgs as depicted in Eq.~\ref{ep} can be naturally realized in a Two Higgs Doublet Model (2HDM) with an approximate $\mathbb{Z}_2$ symmetry, where $H_2$ is odd, and $H_1$ together with the SM fields are even.
In the exact $\mathbb{Z}_2$-symmetry limit, one possible configuration for the vacuum, that respects the symmetry, is $\left<H_1\right>\neq 0$, $\left<H_2\right> = 0$.
In this case, known as the Inert Doublet Model (IDM), the $\mathbb{Z}_2$-odd fields $H^{\pm}$, $A$, and $\phi$ are forbidden to decay to final states with only $\mathbb{Z}_2$ even fields.~\cite{Deshpande:1977rw,Ginzburg:2010wa}
In particular, the lightest $\mathbb{Z}_2$-odd neutral scalar is completely stable and can provide a dark matter candidate \cite{Ma:2006km,Barbieri:2006dq,LopezHonorez:2006gr,Goudelis:2013uca,Arhrib:2013ela,Ilnicka:2015jba,Belyaev:2016lok}.

In the IDM, electroweak symmetry breaking (EWSB) directly contributes to the masses of the $\mathbb{Z}_2$-odd Higgses.
In particular, if we take the limit where the mass term of the inert doublet ($m_2^2H_2^2$) approaches zero, the masses of the $\mathbb{Z}_2$-odd Higgsses are solely generated via EWSB.
In this limit $m^2_{A,\phi,H^{\pm}}\sim \la v^2$, where $\la$ represents the quartic couplings for $\mathbb{Z}_2$-conserving terms such as $|H_1|^2|H_2|^2$.
Therefore, $\mathbb{Z}_2$-preserving quartic couplings between $H_1$ and $H_2$ can be responsible for the mass hierarchy between SM and non-SM Higgs bosons.
This is an example of \emph{non-decoupling} electroweak symmetry breaking and motivates the study of triple-Higgs couplings $g_{hHH}$, where $h$ is the SM Higgs boson and $H=H^{\pm},A,\phi$.

One way to study the coupling $g_{hHH}$ is to look at the decays of the SM-Higgs boson $h$.
If the mass of the lightest $\mathbb Z_2$-odd Higgs $H$ is less than half the mass of the SM Higgs, $h$ can decay to a pair of on-shell $\mathbb Z_2$-odd Higgses. Studies on exotic decays of the SM Higgs \cite{Sirunyan:2018mot,Aaboud:2018esj,Sirunyan:2018owy,Aaboud:2019rtt} largely constrain $g_{hHH}$ in this case.
Here we focus on the scenario where $H$ is heavy, so that exotic $h$ decays cannot probe $g_{hHH}$ efficiently. 
Instead, we study $g_{hHH}$ by pair producing $H$ via an off-shell $h$ at the Large Hadron Collider (LHC).
Depending on which of the additional scalars ($H^{\pm}$, $A$, or $\phi$) is the lightest $\mathbb Z_2$-odd state, different search strategies are required to constrain $g_{hHH}$.
If $H^\pm$ is long-lived, searches looking for heavy stable charged particles (HSCP) \cite{Khachatryan:2016sfv,CMS:2016ybj} can be sensitive.
If the neutral scalar or pseudoscalar is long-lived, searches using missing transverse energy (MET) recoiling against initial state radiation~\cite{Aaboud:2017phn,CMS:2017tbk,Aaboud:2018xdl} can be sensitive.
In this work, we relax the exact $\mathbb{Z}_2$-symmetry condition, so that small explicit $\mathbb Z_2$-breaking terms are allowed. 
With this assumption, the approximately $\mathbb{Z}_2$-odd states can start to decay promptly inside the detector, providing a much richer phenomenology at the LHC \cite{1812.08179}.

In general, $\mathbb{Z}_2$ breaking terms can be introduced in the Higgs potential via odd powers of $H_2$ or in the Yukawa sector through terms proportional to $H_2$.
In Ref.~\cite{1812.08179}, we introduced $\mathbb{Z}_2$ breaking terms in the Higgs potential as well as in the Yukawa sector, so these two $\mathbb{Z}_2$-breaking sources were independent.
In this work, we only consider $\mathbb{Z}_2$ breaking terms in the scalar potential, for the purpose of staying within the context of the type-I 2HDM.
The $\mathbb{Z}_2$ breaking terms will introduce a VEV $\left<H_2\right> \sim \ep v$ and induce $\mathcal{O}(\ep)$ suppressed couplings for interactions between one $\mathbb{Z}_2$-odd field and SM fields.

The smallness of the $\mathbb{Z}_2$-violating couplings will not prevent a sizeable triple Higgs coupling $g_{hHH}$, since the latter is dominated by the $\mathbb{Z}_2$-preserving quartics.
Consequently, as mentioned before, the production mode of interest is di-Higgs production, $pp\to h^*\to HH$, where $H=A,\phi,H^{\pm}$.
Since the coupling between $H$ and SM fermions is suppressed by $\ep$, the box diagram contributing to this process, dominated by a top loop, can be safely ignored.
This leaves di-Higgs production via off-shell SM Higgs as the only important diagram, unlike for SM-Higgs pair production, where the interference between the box and the triangle diagrams is important.
There are many CMS and ATLAS searches \cite{CMS:2018obr} targeting SM di-Higgs production via the final states $b\bar{b}b\bar{b}$ \cite{Aaboud:2018knk,Sirunyan:2018zkk}, $b\bar b\gamma\gamma$ \cite{CMS:2017ihs,Sirunyan:2018iwt,Aaboud:2018ftw}, $b\bar b WW$ \cite{Aad:2015xja,Sirunyan:2017guj}, and $b\bar b \tau \tau$ \cite{Sirunyan:2017djm,Aaboud:2018sfw}, thus constraining the SM triple Higgs coupling $g_{hhh}$.
However, an enhanced $g_{hHH}$ coupling could also be constrained by these searches.
Here we study the sensitivity of these analyses to the non-SM triple Higgs couplings $g_{hHH}$.

Another reason to look for di-Higgs production via off-shell SM Higgs is that this production mode can be a complementary probe to electroweak pair production (EWPP) in searches looking for almost inert Higgs bosons \cite{1812.08179}.
Exotic Higgses produced through EWPP (via $W$ or $Z$) are always distinct pairs of physical states: $\phi H^{\pm}$, $AH^{\pm}$, $\phi A$ or $H^+H^-$.
Therefore, if the neutral scalar $\phi$ is the lightest approximately $\mathbb{Z}_2$-odd state, pairs of $\phi$ cannot be produced directly via electroweak processes.
In this case, $pp\to h^*\to\phi\phi$ is kinematically favored and could become the most important production channel for exotic Higgses.
Since pairs of carged Higgses $H^+H^-$ can be produced through EWPP, in this work we focus on pair production of the neutral states ($\phi$ and $A$).

The paper is organized in the following way.
In Section 2 we present the model, that corresponds to a Type I 2HDM with small $\mathbb Z_2$-breaking terms, and discuss the production and decay modes of the approximately inert Higgses.
In Section 3 and 4 we find current constraints on $g_{h\phi\phi}$ and $g_{hAA}$ coming from SM di-Higgs searches. Section 5 contains our conclusions and outlook.

\section{Almost inert Higgs bosons}
In this section we present the model and introduce the relevant parameters for the phenomenology.
Consider two Higgs doublets that have the following $\mathbb{Z}_2$ symmetry: 
\begin{equation}
H_1 \mapsto H_1,
\qquad
H_2 \mapsto -H_2.
\end{equation}
The most general $\mathbb{Z}_2$-conserving potential is
\begin{equation}\label{potential}
\begin{split}
V_0 =\,&m^2_1 H_1^{\dagger}H_1 + \frac{1}{2}\la_1 (H_1^{\dagger}H_1)^2
+m^2_2 H_2^{\dagger}H_2+\frac{1}{2}\la_2 (H_2^{\dagger}H_2)^2\\
&
+\la_3 (H_2^{\dagger}H_2)(H_1^{\dagger}H_1)
+\la_4 (H_2^{\dagger}H_1)(H_1^{\dagger}H_2)
+\frac{1}{2}[\la_5 (H_2^{\dagger}H_1)^2+h.c.].
\end{split}
\end{equation}
We assume that $m_1^2<0$ and $m_2^2>0$, such that in the exact $\mathbb{Z}_2$-symmetry limit, only $H_1$ gets a VEV.
It is instructive to trade the parameters in the Higgs potential with the phenomenological parameters. 
Assuming all the parameters are real (so there is no explicit CP violation), we trade the parameters in the potential for the following parameters:
\[
v,m_h,m_{\phi},m_A,m_{H^\pm},\la_2,m_2^2,\label{pheno_param}
\]
where $v$ and $m_h$ are the VEV and mass of the SM Higgs (which resides in $H_1$), while $m_{\phi}$, $m_A$, and $m_{H^+}$ are the masses of the additional neutral scalar, pseudoscalar and charged scalar, all of which reside in the inert doublet $H_2$.
This vacuum configuration does not break the symmetry $H_2\to -H_2$, so the lightest $\mathbb Z_2$-odd particles are stable. This is also known as the Inert Doublet Model (IDM).

A large mass splitting between $m_{H^{\pm}}$ and $m_A$ would result in large violations of the $SU(2)$ custodial symmetry, which is very well constrained by electroweak precision tests.
Therefore, in this work we assume $m_A=m_{H^{\pm}}$.

\subsection{Di-Higgs production at the LHC}

\begin{figure}[h]
\begin{center}
\begin{subfigure}[b]{0.47\textwidth}
\includegraphics[width=\textwidth]{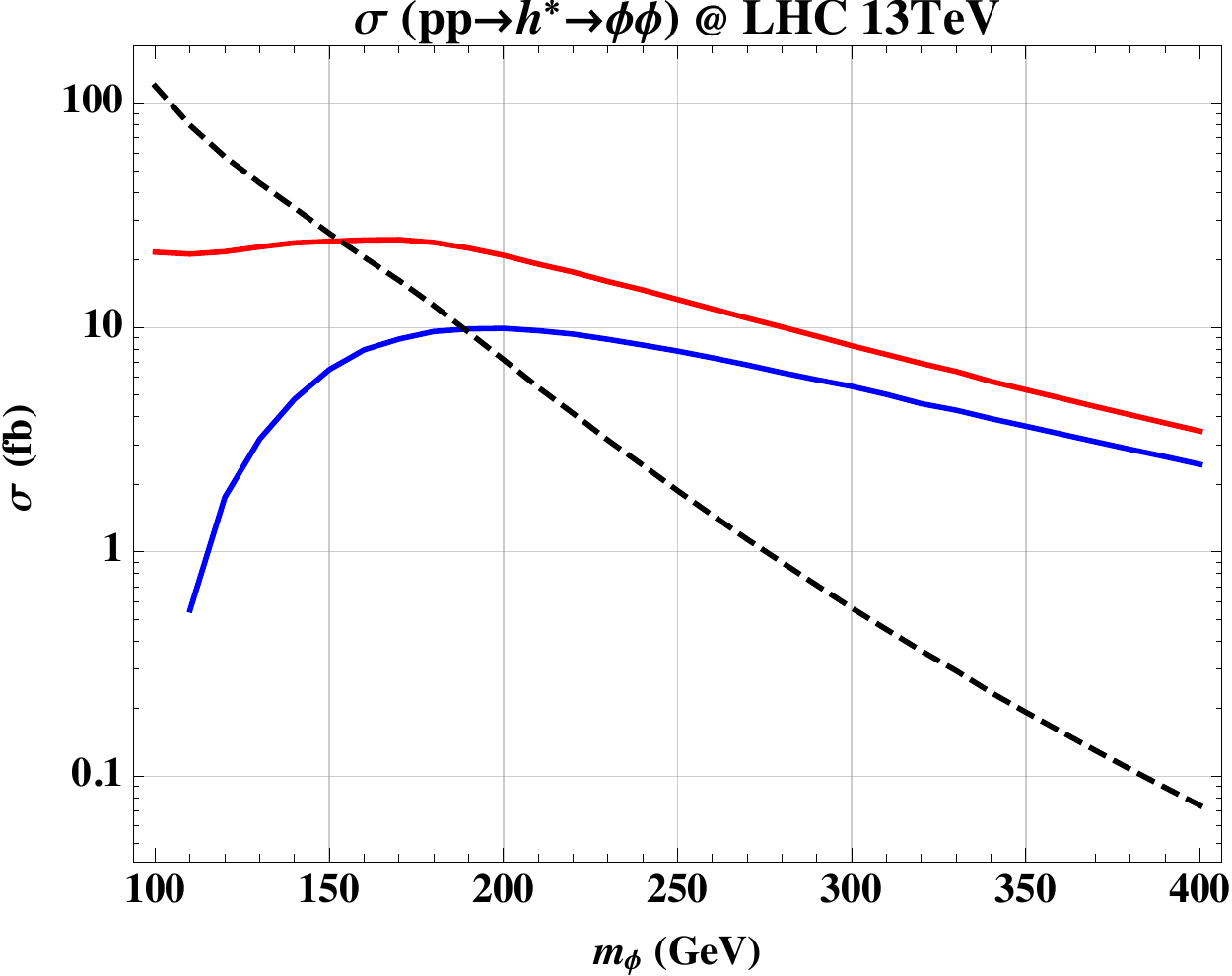}
\end{subfigure}
\hfill
\begin{subfigure}[b]{0.485\textwidth}
\includegraphics[width=\textwidth]{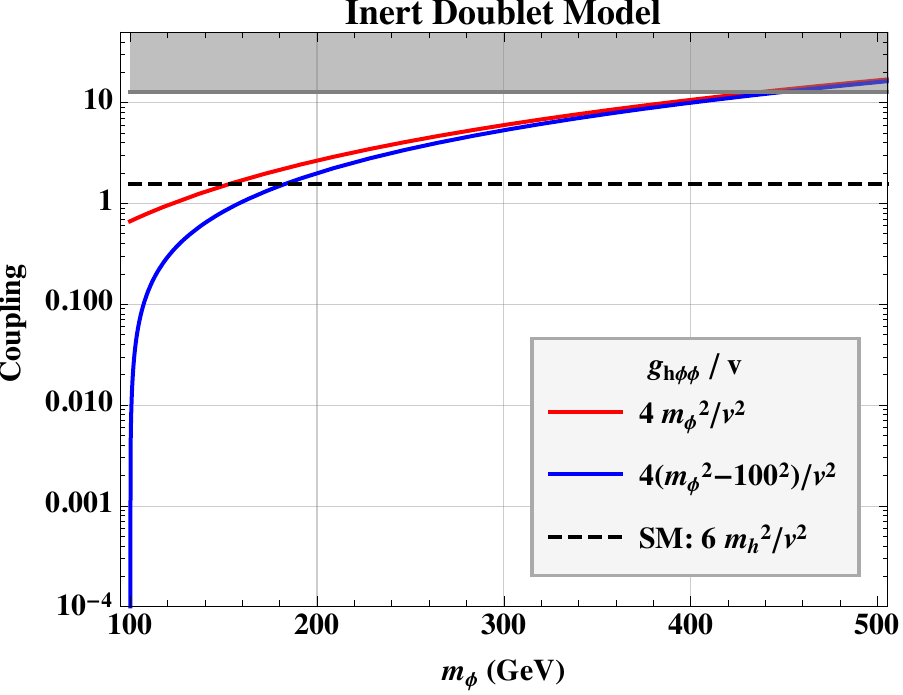}
\end{subfigure}
\caption{Left panel shows examples of the production cross section of $gg\to h^*\to \phi\phi$. Differently colored curves assume different choices of couplings, which are shown in the right panel.
The gray region indicates the strong coupling regime where the theory loses perturbativity.
Analytical expressions are taken from Ref.~\cite{hep-ph/9603205}.}\label{xsec}
\end{center}
\end{figure}

We investigate the non-SM di-Higgs production cross section, \emph{i.e.} $gg\to h^*\to \phi\phi (AA)$, in the exact $\mathbb{Z}_2$-symmetry limit (IDM).
First, we extract the triple-Higgs coupling $g_{hHH}$, where $H=\phi,A,H^+$.
As discussed in the introduction, these couplings arise from $\mathbb{Z}_2$-preserving quartic terms in the Higgs potential:
\begin{equation}
\begin{split}
&g_{hH^+H^-}=v\la_3,
\\
&g_{hAA}=2v(\la_3+\la_4-\la_5),
\\
&g_{h\phi\phi}=2
v(\la_3+\la_4+\la_5),
\end{split}
\end{equation}
where the factor of 2 in $g_{hAA}$ and $g_{h\phi\phi}$ will compensate the symmetry factor in the relevant terms of the Lagrangian, where identical particles appear\footnote{Similarly, the SM triple Higgs coupling $g_{hhh}$ will be $3\la_1 v$.
}.
To see what range of values $\la_3$, $\la_4$, and $\la_5$ can take, it is instructive to rewrite them as:
\[
\begin{split}
&\la_3 v^2= 2(m_{H^+}^2-m_2^2),
\\
&\la_4v^2=\la_5v^2+2(m^2_A-m^2_{H^{\pm}}),
\\
&\la_5v^2=m^2_{\phi}-m^2_A.\label{eq:la_masses}
\end{split}
\]
Therefore, in the IDM we have:
\begin{equation}
\begin{split}
g_{hH^+H^-} &= 2(m_{H^+}^2-m_2^2)/v,\\
g_{hAA}     &=4(m_{A}^2-m_2^2)/v,\\
g_{h\phi\phi} &=4(m_{\phi}^2-m_2^2)/v,
\end{split}
\label{gphiphi}
\end{equation}
where $m_2^2>0$.
In the case where $m_2^2=0$, the masses of the exotic Higgses are solely generated by the $\mathbb Z_2$-preserving quartic couplings $\lambda_{3,4,5}$ upon EWSB. 
In particular, $\lambda_3$ scales with the mass squared of the additional scalars.
We can see that even a mild mass hierarchy $m_{\phi,A,H^{\pm}}\gtrsim m_h$ produces $\lambda_3 \sim \mathcal O(1) \gg \lambda_1$, and consequently a sizeable di-Higgs production rate.
This will partially compensate for the drop in the gluon parton distribution function in the large mass regime. 
As an example, Fig.~\ref{xsec} shows the cross section of $gg\to h^*\to \phi\phi$ for $m_2^2=0$ (red solid) and $m_2^2=(100\mbox{ GeV})^2$ (blue solid).
One can see that for different choices of $m_2^2$, the cross section can change for orders of magnitude at a given mass. 
However, the coupling soon reaches 4$\pi$ when $m_{\phi} \gtrsim 400$ GeV. Therefore, the theory reaches a non-perturbative regime and a leading order estimate of the cross section is no longer reliable.
For this reason, we restrict ourselves to 
\[
125\, \mbox{GeV} <m_{\phi}<400\,\mbox{GeV}.
\]
For comparison, in Fig.~\ref{xsec} we also plot the cross section with the coupling set to be a constant equal to the SM triple Higgs coupling (black dashed).
Its steep fall is due to the fast drop in the gluon parton distribution function.

\subsection{Decay of approximately inert Higgs}\label{z2breaking}
In the IDM, the lightest $\mathbb{Z}_2$-odd neutral state can only appear as missing energy at the LHC. However, if the symmetry is not exact, it can decay within the detector, which is what we investigate in this section.

Starting with the Higgs potential, Eq.~\ref{potential}, one can introduce $\mathbb{Z}_2$ breaking via including odd powers of $H_1$:
\begin{equation}
\Delta V=\Delta m^2(H_2^\dagger H_1+h.c.)
+\Delta\la(H_2^{\dagger}H_2)(H_2^{\dagger}H_1+h.c.)+\Delta\la'(H_1^{\dagger}H_1)(H_2^{\dagger}H_1+h.c.),
\end{equation}
where the $\mathbb{Z}_2$-breaking terms are all of $\mathcal{O}(\ep) \ll 1$.
Now the `inert' doublet $H_2$ gets a VEV of $\mathcal{O}(\ep)$:
\begin{equation}\label{eq:v2}
\langle H_2\rangle
\equiv \tilde{v}_2= -\frac{\Delta m^2/v^2+\Delta\la/2}{m_2^2/v^2+(\la_3+\la_4+\la_5)/2} v +\mathcal{O}(\ep^2).
\end{equation}
The ratio of the VEVs of the two doublets defines the conventionally adopted parameter $\tan\beta$ in 2HDM. In the approximate $\mathbb{Z}_2$ limit, it is given by $v/\tilde{v}_2$, and tend to be large.\footnote{In the almost inert limit, $H_{1,2}\approx \Phi_{2,1}$, where $\Phi$ represents the usually defined $\mathbb Z_2$ basis. Consequently, in this limit, $\tan\beta \equiv v_2/v_1\approx \tilde v_1/\tilde v_2 \approx v/\tilde v_2$, where $v_{1,2}\equiv \left<\Phi_{1,2}\right>$.
See the appendix of Ref.~\cite{1812.08179} for more details.}

The fields can be parametrized around their VEVs as
\begin{equation}\label{pars}
H_i=
\left (
  \begin{array}{c}
  H_i^+ \\
 \frac{1}{\sqrt{2}}(\tilde v_i+h_i^0+iA_i)
  \end{array}
\right ).
\end{equation}
All the physical fields receive an $\mathcal{O}(\ep)$ correction with respecto to their $\mathbb Z_2$-symmetric limit.
The neutral pseudoscalar and charged components of the doublets mix to generate the physical fields $A$ and $H^{\pm}$:
\begin{equation}
\begin{split}
A  &= A_2 + \epsilon_A A_1 +\mathcal{O}(\ep^2),\\
H^+  &= H^+_2 + \epsilon_A H^+_1 +\mathcal{O}(\ep^2),
\end{split}
\end{equation}
where
\begin{equation}\label{eq:epA}
    \epsilon_A = -\frac{\tilde v_2}{v} = \mathcal{O}(\ep).
\end{equation}
Note that $\epsilon_A$ is related to the usual $\tan\beta$ parameter through $\epsilon_A \approx -1/\tan\beta$. The physical scalar fields are
\begin{equation}
\begin{split}
\left(
 \begin{array}{c}
h\\
\phi
  \end{array}
\right)&=
\left(
 \begin{array}{cc}
1 & \ep_h\\
- \ep_h & 1\\
  \end{array}
\right)
\left(
 \begin{array}{c}
H_1^0\\
H_2^0
  \end{array}
\right) +\mathcal{O}(\ep^2),
\end{split}
\end{equation}
with
\begin{equation}
\ep_h=\frac{1}{m_h^2-m_{\phi}^2}\Big{[}\frac{\tilde v_2}{v}(m_{\phi}^2-2m_{H^{\pm}}^2+\la_3 v^2)
+\Delta\la v^2
\Big{]} = O(\ep).
\end{equation}
Note that $\ep_h$ is equivalent to $\alpha$ ($\alpha-\pi/2$) when $m_h<m_{\phi}$ ($m_h>m_{\phi}$), where $\alpha$ is another conventionally adopted parameter in 2HDM.
From the kinetic terms, we find how $\mathbb{Z}_2$-odd fields couple to SM fields:
\[
g_{\phi VV}\propto \ep_V,\quad g_{A Zh}\propto \ep_V,\quad g_{H^{\pm} W^{\mp}h}\propto \ep_V,
\]
where 
\begin{equation}
\ep_V\equiv \epsilon_A + \ep_h.    
\end{equation}
The limit $\ep_V\to 0$ corresponds to the so-called \emph{alignment limit} in 2HDM \cite{Gunion:2002zf,Craig:2013hca,Carena:2013ooa,Delgado:2013zfa}.

\begin{figure}
\begin{center}
\begin{subfigure}[b]{0.43\textwidth}
\includegraphics[width=\textwidth]{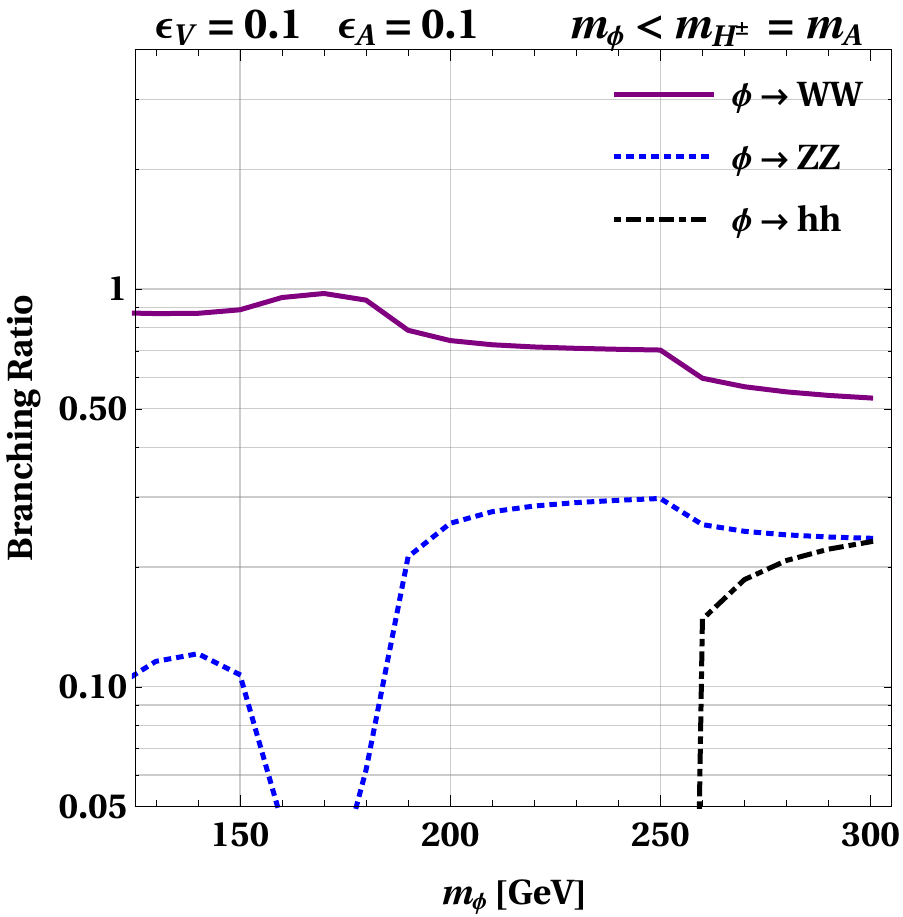}
\caption{}\label{phibr_WW}
\end{subfigure}
\hfill
\begin{subfigure}[b]{0.43\textwidth}
\includegraphics[width=\textwidth]{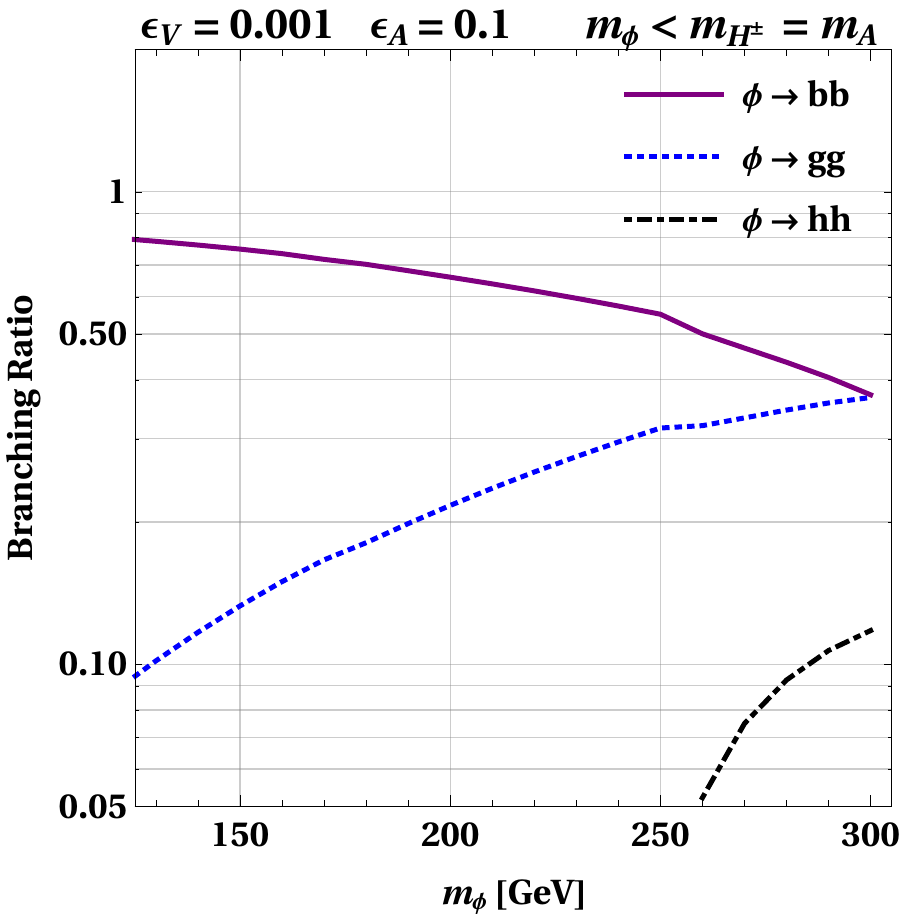}
\caption{}\label{phibr_bb}
\end{subfigure}
\hfill
\begin{subfigure}[b]{0.43\textwidth}
\includegraphics[width=\textwidth]{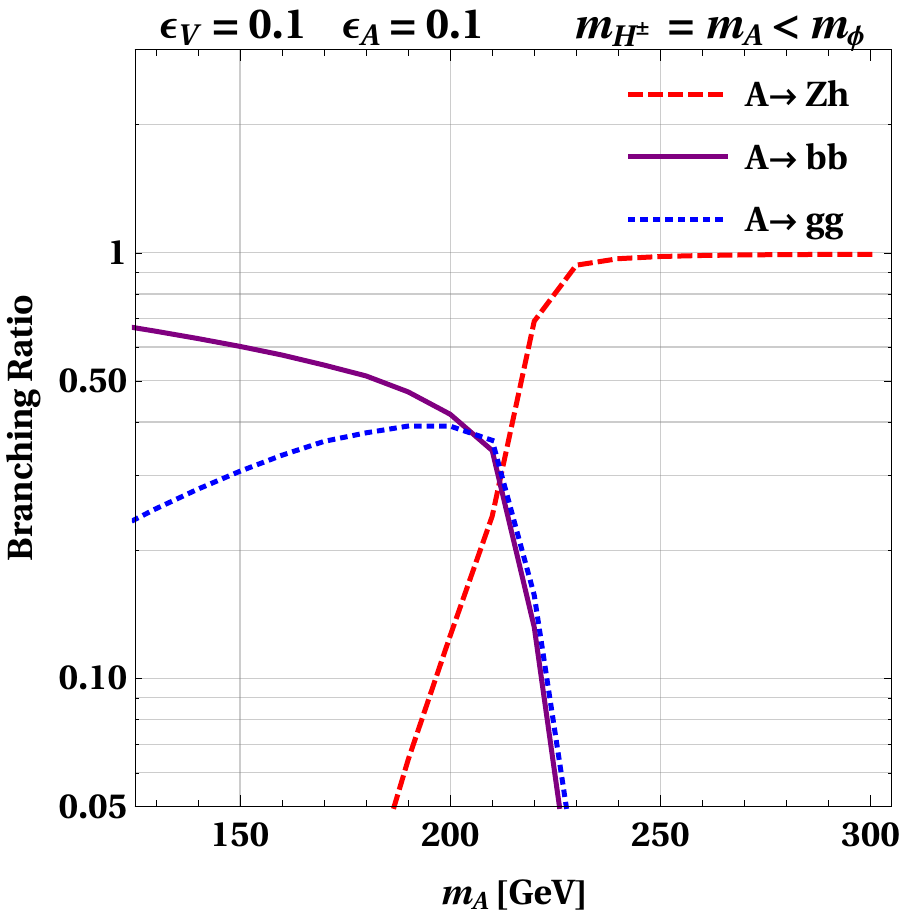}
\caption{}\label{Abr}
\end{subfigure}
\hfill
\begin{subfigure}[b]{0.43\textwidth}
\includegraphics[width=\textwidth]{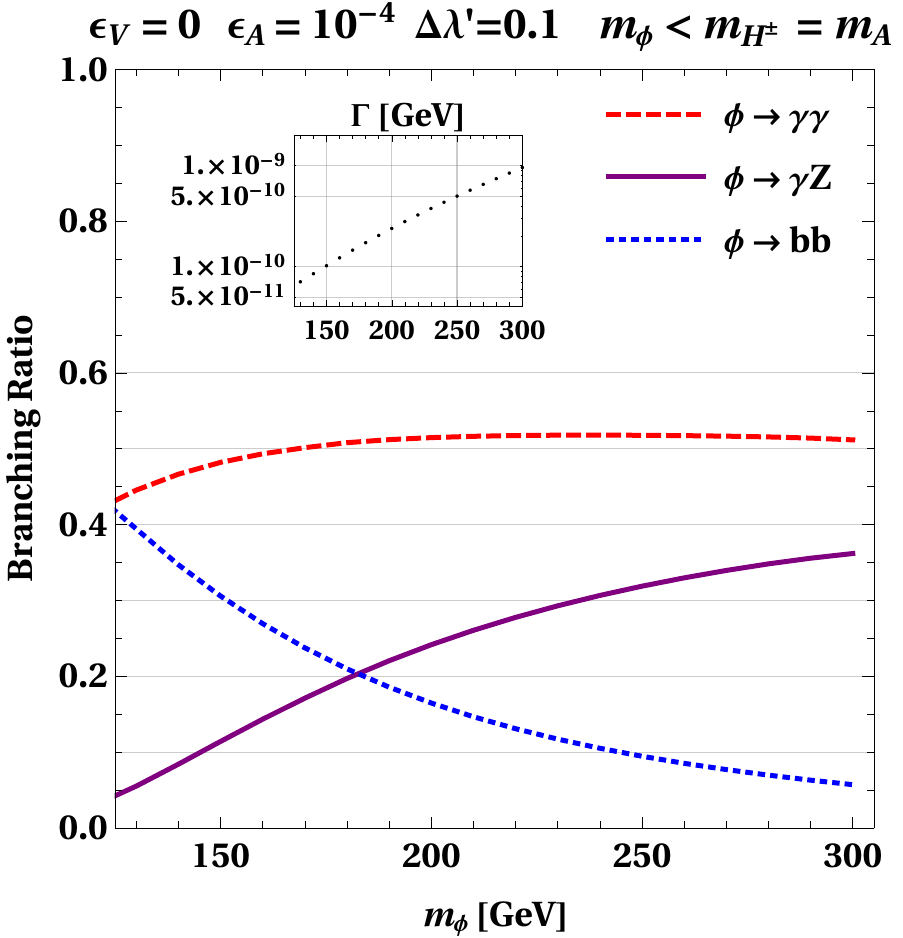}
\caption{}\label{phibr_photon}
\end{subfigure}
\caption{Examples of dominant decay modes of the neutral approximately $\mathbb{Z}_2$-odd Higgsses as a function of their mass.
The small $\mathbb{Z}_2$ breaking parameters $\ep_V$ and $\ep_A$ roughly control the size of their decays to SM weak gauge bosons and fermions, respectively. 
The subplot in Fig.~\ref{phibr_photon} shows that $\phi$ can still undergo a prompt decay even when $\ep_V\sim\ep_A\to 0$, via loop induced processes which are controlled by $\Delta\la'$. }\label{plotphibr}
\end{center}
\end{figure}

The $\mathbb{Z}_2$-breaking terms in the Higgs potential also induce $\mathcal{O}(\ep)$ $\mathbb{Z}_2$-violating terms in the Yukawa sector.
In the exact $\mathbb{Z}_2$-symmetry limit, only $H_1$ can have Yukawa couplings, therefore, this corresponds to a Type I 2HDM.
After $\mathbb{Z}_2$ breaking, the approximate $\mathbb{Z}_2$-odd scalars couple to SM fermions via an $\mathcal{O}(\ep)$ VEV and/or mass mixing:
\[
g_{\phi f \bar{f}}\propto \ep_V-\epsilon_A,
\quad g_{Af\bar{f}}\propto \epsilon_A,
\quad  g_{H^{\pm}f\bar{f'}}\propto \epsilon_A.
\]
In summary, the following set of phenomenological parameters can be used to discuss the approximate $\mathbb{Z}_2$ Higgses:
\begin{equation}
v,m_h,m_{\phi},m_A,m_{H^{\pm}}, m_2^2, \ep_V,\epsilon_A,
\end{equation}
where we have left out $\la_2$ and $\Delta\la'$.
Both terms have more than two powers of $H_2$. Since $\langle H_2 \rangle=\tilde{v}_2 \sim \mathcal{O}(\ep)$, these terms are not important at the tree level.
Fig.~\ref{plotphibr} shows examples of $\phi$ branching ratios.
It can be seen that adjusting the relative size of $\ep_V$ to $\epsilon_A$ radically changes $\phi$'s dominant decays.

\subsubsection*{A special case: $\phi\to \gamma\gamma$ }
If $\phi$ is the lightest approximate $\mathbb{Z}_2$-odd Higgs, in the limit that $\ep_V\to 0$, $\epsilon_A \to 0$, we expect that $\phi$ will never decay inside the detector.
However, in this scenario, where the tree-level decays of $\phi$ are suppressed, one can no longer ignore loop induced processes. In particular, the quartic interaction $\Delta\la'|H_2|^2(H_1H_2^{\dagger}+h.c.)$ contributes a $\mathbb{Z}_2$-violating coupling $g_{\phi H^+H^-}\sim \Delta\la' v$.
In the limit that $\Delta\la'\gg \ep_V\sim \epsilon_A \to 0$, the dominant decay of $\phi$ can become $\gamma\gamma$/$\gamma Z$ induced by an $H^+$ loop, with one caveat.
The inclusion of a term like $\Delta\la'|H_2|^2(H_1H_2^{\dagger}+h.c.)$ necessarily introduces the $\mathbb{Z}_2$-breaking term $\Delta m^2 H_1^{\dagger}H_2$ back at one loop.
If $\Lambda$ is the UV cutoff of the model, $\Delta m^2 $ can be estimated to be $\sim\frac{\Delta\la'}{16\pi^2}\Lambda^2$. From Eqs.~\ref{eq:v2} and \ref{eq:epA}, we can see that the limit $\Delta\la'\gg \ep_V\sim \epsilon_A \to 0$ is only achievable if $\Lambda\lesssim 4\pi v \sim 3$ TeV.

Fig.~\ref{phibr_photon} shows an example of the branching ratios of $\phi$ in this region of the parameter space.
In the subplot, we also show the total width of $\phi$.
Therefore, even in the limit that $\ep_V\sim \epsilon_A \to 0$, $\phi$ can still decay within the detector. More importantly, it decays with a spectacular signal involving multiple photons, which have much smaller SM backgrounds.

\section{Constraining $h\phi\phi$ coupling}

In this section, we begin detailed benchmark studies under the assumption of different schemes of $\mathbb{Z}_2$-breaking parameters, as studied in Section~\ref{z2breaking}.
Even though all the $\mathbb{Z}_2$-breaking parameters are assumed to be small, the main decay modes of $\phi$ can change drastically depending on the relative size between them (cf. Fig.~\ref{plotphibr}).
Since there are no a priori assumptions on the origin of the $\mathbb{Z}_2$ breaking, we have to study all possible qualitatively-different decays of $\phi$ at the LHC.
Fig.~\ref{plotphibr} shows the main decay modes for different regions of the parameter space.
Each of the following subsections focus on a scenario where $\phi$ dominantly decays to only one or two types of SM particles, based on which, limits of $g_{h\phi\phi}$ as a function of mass are drawn.
If the LHC is insensitive to such ideal scenarios, we may conclude that it is not going to be sensitive to $g_{h\phi\phi}$ in a more general scheme.

For all the following studies, we generate the parton level events for signal with MadGraph5 \cite{Alwall:2011uj}, followed by hadronization and showering by Pythia8 \cite{Sjostrand:2007gs}, and detector simulation with Delphes3 \cite{Ovyn:2009tx}.
Each subsection assumes a distinct relative size between $\ep_V$ and $\epsilon_A$ and discusses possible constraints on $g_{h\phi\phi}$ arising from the dominant decays.

\subsection{$\ep_A\gg \ep_V: \phi\phi\to(b\bar{b})(b\bar{b})$}\label{sec:4b}

As shown in Fig.~\ref{phibr_bb}, just like the SM Higgs, $\phi$ can dominantly decay to a pair of $b$ quarks.
The signature of interest here is four $b$ jets.
Since both ATLAS and CMS have performed SM di-Higgs searches in this channel \cite{Aaboud:2018knk,CMS:2018smw}, our goal is to see whether these standard searches can be sensitive to the almost inert heavy Higgses.
We follow the selection and cuts from the ATLAS di-Higgs search at $36 \mbox{ fb}^{-1}$ \cite{Aaboud:2018knk}, excepting one cut involving the b-tagging score.
The interested reader can find the details of these cuts in Appendix~\ref{app:bbbb}.
The right panel of Fig.~\ref{limit4b} shows the efficiencies of both signal reconstruction and final selection, where reconstruction simply requires 4 $b$-tagged jets with $p_T>40\mbox{ GeV}$.
\begin{figure}[h]
\begin{center}
\begin{subfigure}[b]{0.48\textwidth}
\includegraphics[width=\textwidth]{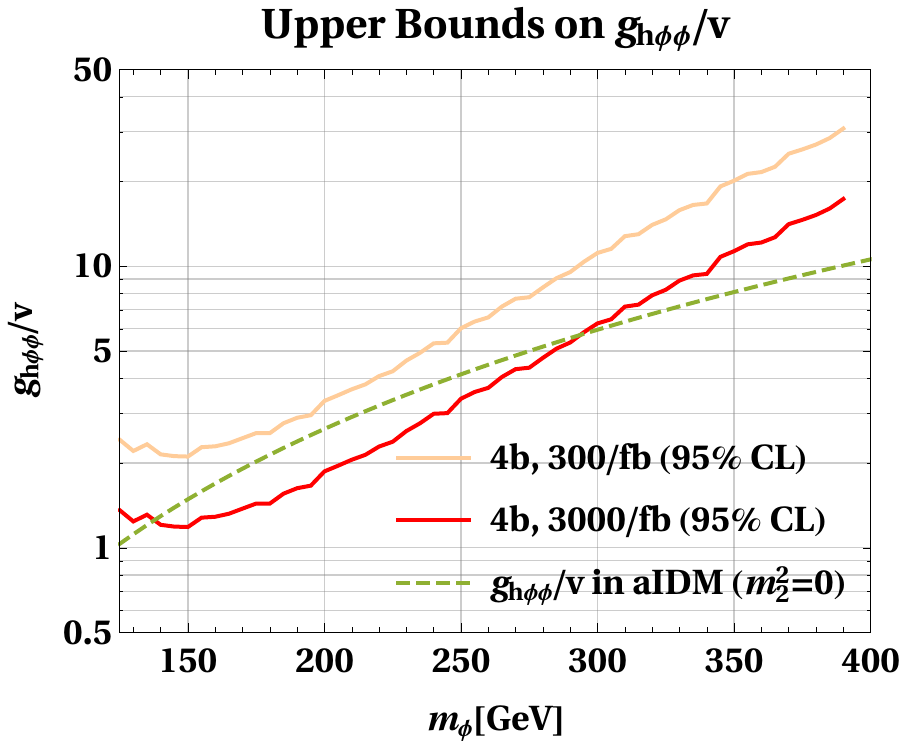}
\end{subfigure}
\hfill
\begin{subfigure}[b]{0.47\textwidth}
\includegraphics[width=\textwidth]{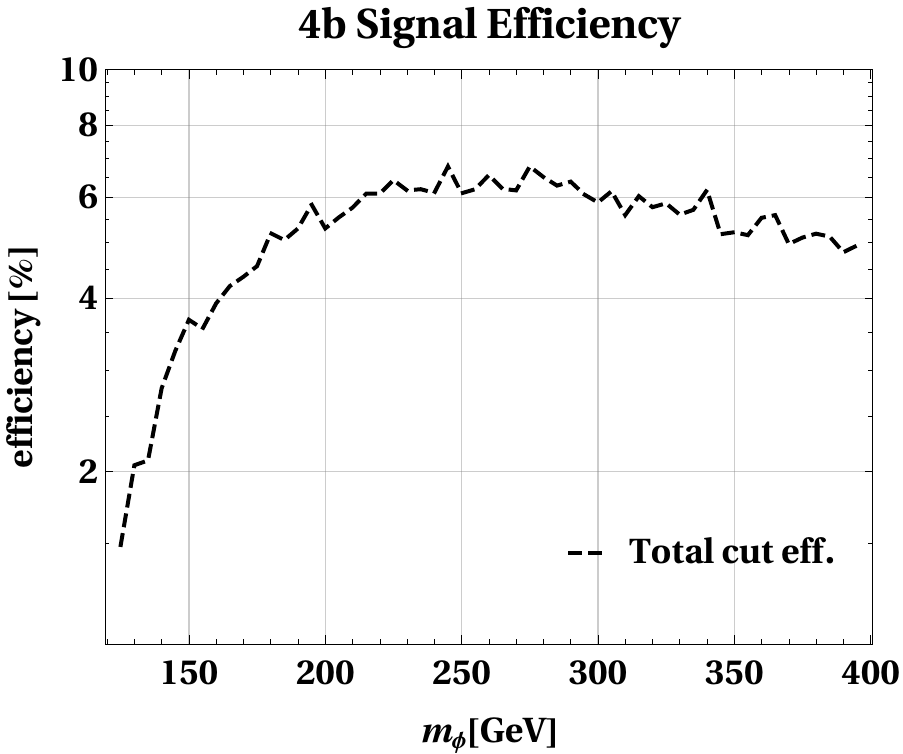}
\end{subfigure}
\caption{Left panel shows the upper limits on $g_{h\phi\phi}/v$ from $pp\rightarrow h \rightarrow \phi\phi\rightarrow (b\bar b)(b\bar b)$ at the LHC with $300\mbox{ fb}^{-1}$ and HL-LHC ($3000\mbox{ fb}^{-1}$). We also include the value of $g_{h\phi\phi}/v$ in aIDM for $m_2^2=0$, which is given by $4m_{\phi}^2/v^2$. Right panel shows the total cut efficiency for the signal, following the analysis from Ref.~\cite{Aaboud:2018knk}.}\label{limit4b}
\end{center}
\end{figure}

The SM backgrounds for this channel are multijets, hadronic $t\bar t$ and semi-leptonic $t\bar t$. 
Both CMS and ATLAS use mainly data-driven methods to model them.
Therefore, to draw limits on coupling vs. mass for higher luminosities, we simply extrapolate the backgrounds at $36\mbox{ fb}^{-1}$ from Ref.~\cite{Aaboud:2018knk}. 
This is a conservative estimate because a real analysis is expected to be more efficient reducing backgrounds for higher values of $m_\phi$.
It is clear from Fig.~\ref{limit4b} that the $4b$ channel of the SM di-Higgs searches can be important in constraining $g_{h\phi\phi}$, even though they are not optimised to target an exotic Higgs.

\subsection{$\ep_A\lesssim \ep_V:\phi\phi\to(WW)(WW)\to (\ell\nu\ell\nu)(\ell\nu jj)$}

As shown in Section \ref{z2breaking}, the coupling of $\phi$ to SM fermions, $g_{\phi f\bar f}$, is proportional to the difference between $\ep_V$ and $\ep_A$.
Since these two parameters are independent, there is a scenario (Fig.~\ref{phibr_WW}) where they cancel each other out, leaving $\phi$ mainly decaying to $W^+W^-$.
Out of all the possibilities for the $4W$ decays, we focus on the final state with 3 leptons and 2 jets.
This is a relatively clean signature, where the additional jets allow triggering.
However, one cannot easily reconstruct the decay chain, therefore this search reduces to lepton counting, which can be potentially covered by several multi-lepton searches at ATLAS and CMS, for example \cite{Aad:2019lpq,Aaboud:2017dmy}.
Our goal here is to investigate whether these standard multi-lepton searches can be sensitive to the almost inert Higgs $\phi$.
We follow the $36\mbox{ fb}^{-1}$ three lepton search adopted in \cite{Sirunyan:2017hvp}, for which details of the cuts are spelled out in Appendix \ref{app:multil}.

\begin{figure}
\begin{center}
\begin{subfigure}[b]{0.48\textwidth}
\includegraphics[width=\textwidth]{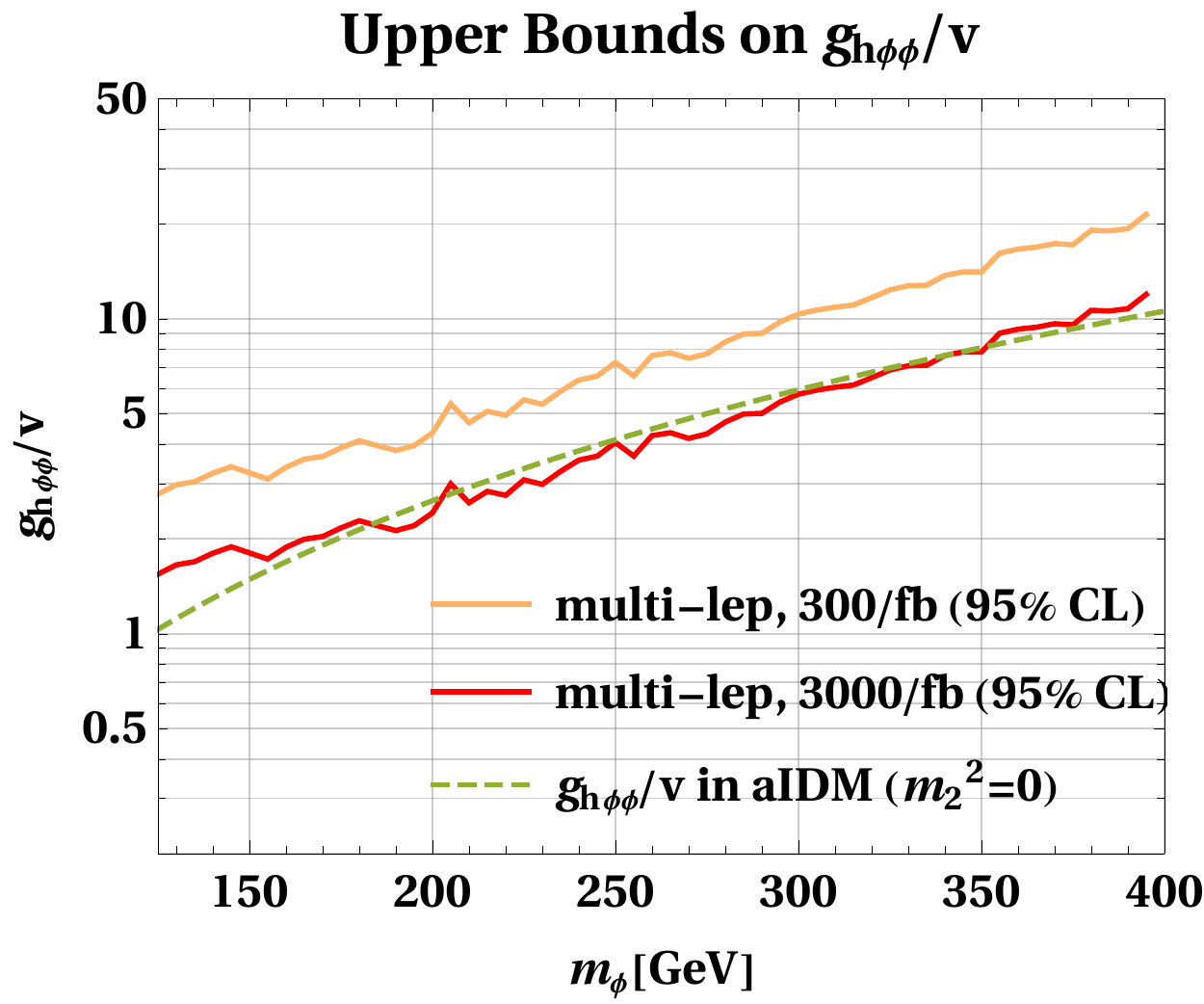}
\end{subfigure}
\hfill
\begin{subfigure}[b]{0.47\textwidth}
\includegraphics[width=\textwidth]{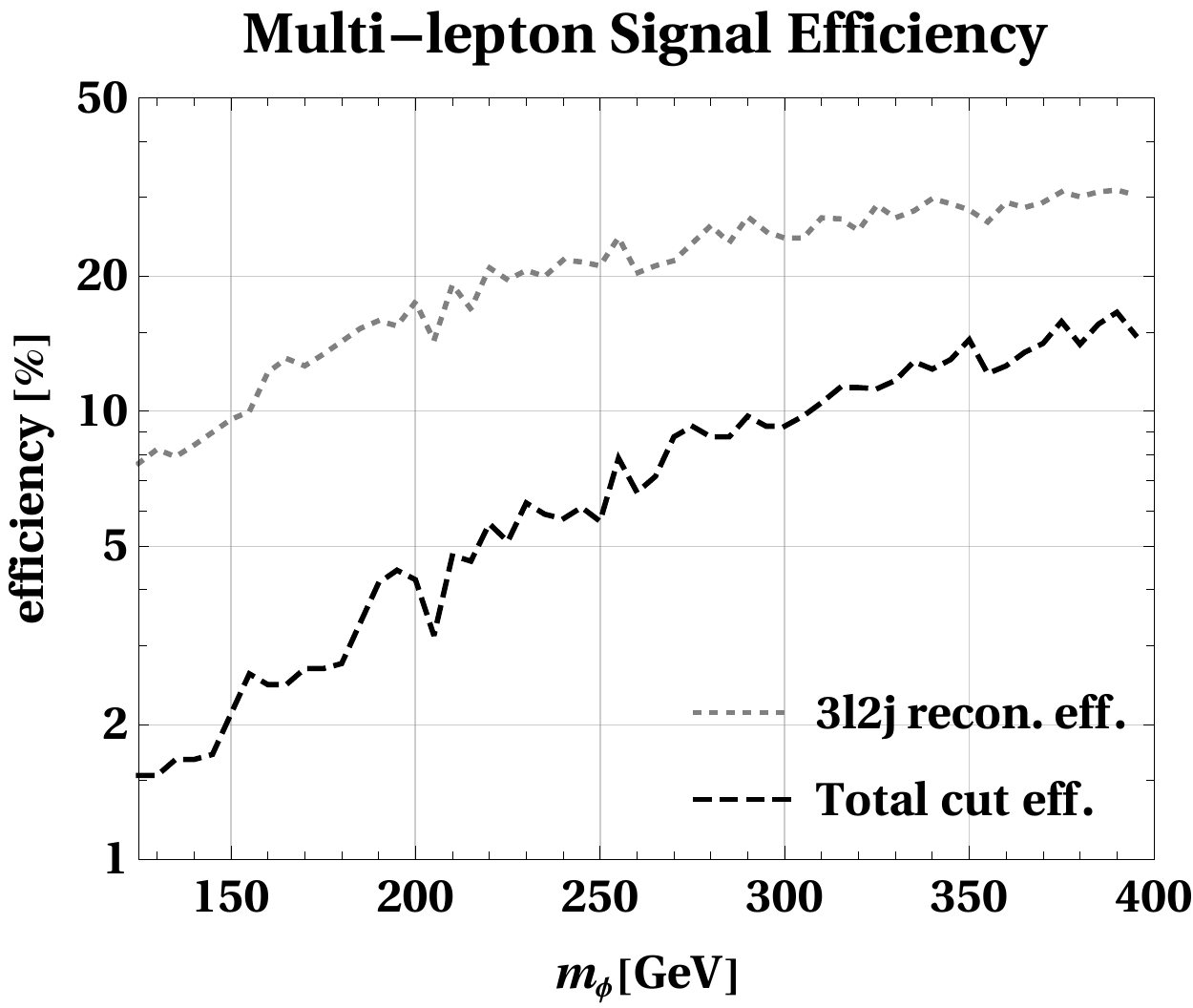}
\end{subfigure}
\caption{Left panel shows the upper limits on $g_{h\phi\phi}/v$ from $pp\rightarrow h \rightarrow \phi\phi\rightarrow (WW)(WW)\rightarrow(\ell\nu\ell\nu)(\ell\nu jj)$ at the LHC with $300\mbox{ fb}^{-1}$ and HL-LHC ($3000\mbox{ fb}^{-1}$). We also include the value of $g_{h\phi\phi}/v$ in aIDM for $m_2^2=0$, which is given by $4m_{\phi}^2/v^2$. Right panel shows the cut efficiency for the signals.}\label{fig:multiL}
\end{center}
\end{figure}

For this channel, the SM backgrounds are mainly non-prompt leptons, diboson and $t\bar{t}V$.
Again, to draw 95\% CL contours on coupling vs. mass for higher luminosities, we simply extrapolate the backgrounds at 36 fb$^{-1}$ from Ref.~\cite{Sirunyan:2017hvp}.
As can be seen in Fig.~\ref{fig:multiL}, this search starts being sensitive to $g_{h\phi\phi}$ only at the high luminosity LHC.

\subsection{$\Delta\lambda'\gg \epsilon_V \sim \epsilon_A:\phi\phi\to(b\bar{b})(\gamma\gamma)$}

In the limit where $\Delta\lambda'\gg \epsilon_A\sim \epsilon_V$, tree level decays of $\phi$ are suppressed, so loop-mediated decays become important.
The interaction $\Delta\lambda'|H_2|^2(H_1^\dag H_2+h.c.)$ contributes to the $\mathbb{Z}_2$-violating vertex $g_{\phi H^+ H^-}\sim v \Delta \lambda'$, allowing the decays $\phi\to \gamma\gamma$ and $\phi\to Z\gamma$.
Therefore, the final states $(\gamma\gamma)(\gamma\gamma)$ and $(\gamma\gamma)(Z \gamma)$ could be sensitive in this scenario, but the analysis would require an estimation of the backgrounds due to fake photons from multijets, which is beyond the scope of this work.
Additionally, if $\epsilon_A$ is small, but not zero, we could also have a sizeable decay to $b\bar b$, as shown in Fig.~\ref{phibr_photon}.
In this scenario, di-Higgs searches in the $b\bar b \gamma\gamma$ final state~\cite{ATL-PHYS-PUB-2017-001,CMS:2017ihs} could be relevant.
On the other hand, if $\epsilon_A\sim\epsilon_V\to 0$, depending on the value of $\Delta\lambda'$, the exotic Higgs could be long lived.
In the latter case, delayed photon searches \cite{Aad:2014gfa,Sirunyan:2019wau} could be sensitive, but these analyses cannot be straightforwardly recast, and a dedicated analysis, beyond the scope of this work, would be required.
Therefore, for this scenario we focus on the prompt $b\bar b\gamma\gamma$ final state.

In particular, we study the potential sensitivity to $g_{h\phi\phi}$ of current ATLAS and CMS di-Higgs searches, in the $b\bar b\gamma \gamma$ final state \cite{ATL-PHYS-PUB-2017-001,CMS:2017ihs}.
For concreteness, we consider the scenario depicted in Fig.~\ref{phibr_photon}, where $\Delta\lambda'=0.1$, $\epsilon_V=0$ and $\epsilon_A = 10^{-4}$.

\begin{figure}[h!]
\begin{center}
\begin{subfigure}[b]{0.48\textwidth}
\includegraphics[width=\textwidth]{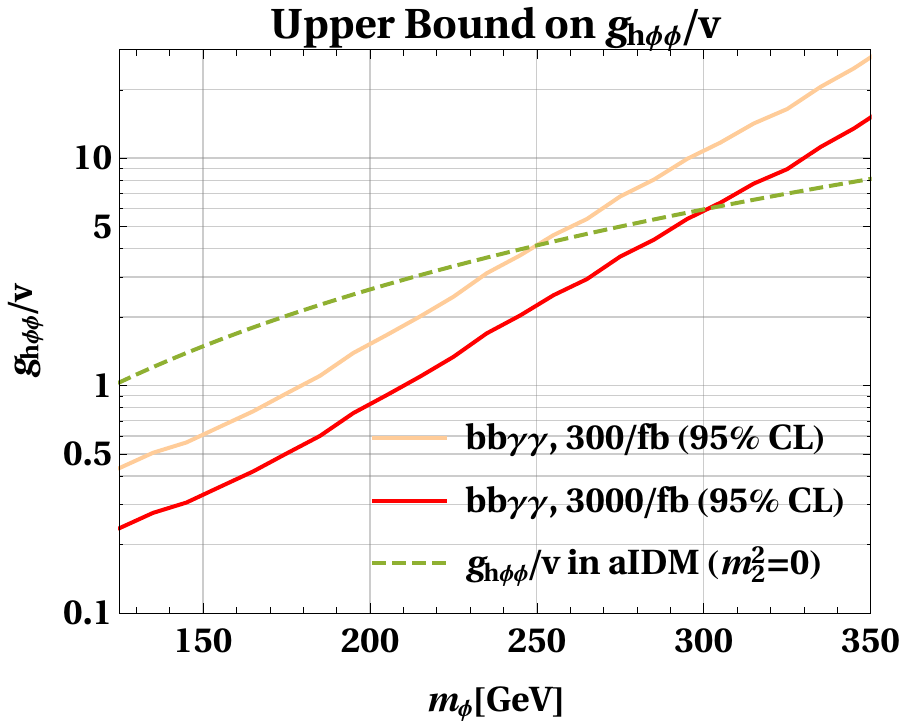}
\end{subfigure}
\hfill
\begin{subfigure}[b]{0.47\textwidth}
\includegraphics[width=0.97\textwidth]{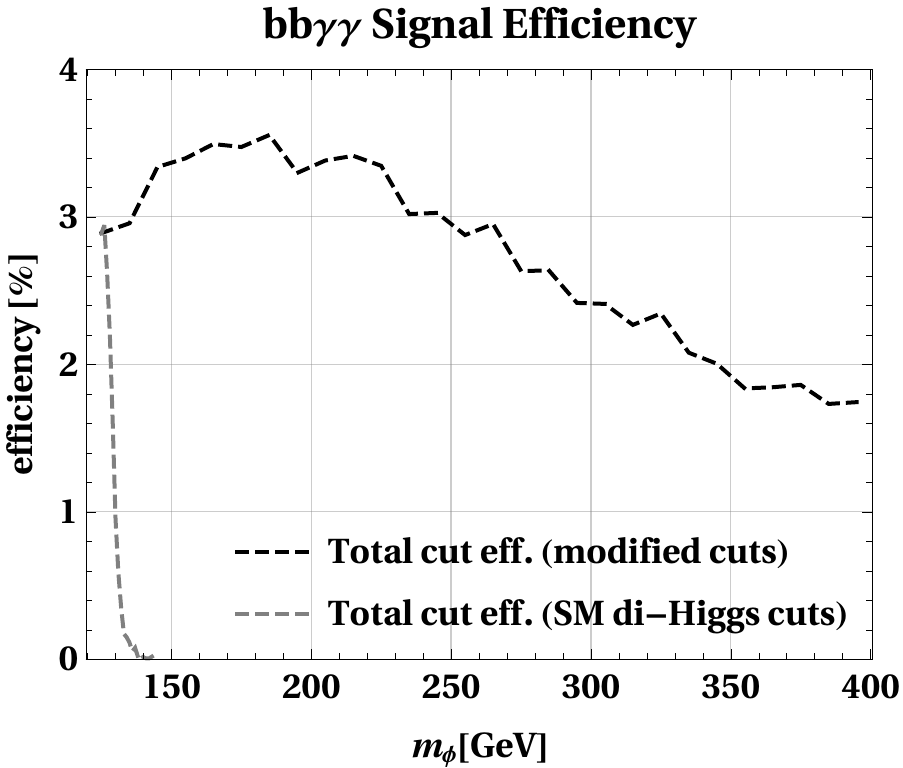}
\end{subfigure}
\caption{Left panel shows the upper limits on $g_{h\phi\phi}/v$ from $pp\rightarrow h \rightarrow \phi\phi\rightarrow (b\bar b)(\gamma\gamma)$ at the LHC with $300\mbox{ fb}^{-1}$ and HL-LHC ($3000\mbox{ fb}^{-1}$).
We also include the value of $g_{h\phi\phi}/v$ in aIDM for $m_2^2=0$, which is given by $4m_{\phi}^2/v^2$. Right panel shows the signal efficiency with the cuts from Ref.~\cite{ATL-PHYS-PUB-2017-001} and using $m_\phi$-dependent invariant mass cuts. }\label{bbgaga_eff}
\end{center}
\end{figure}

We perform a Monte Carlo simulation of the signal, for different masses of the neutral scalar $\phi$, and apply the cuts from Ref.~\cite{ATL-PHYS-PUB-2017-001}, excepting for a modification of the invariant mass cuts.
Contrary to the cuts for the $b\bar b b \bar b$ final state in the previous section, these cuts are optimized for the mass of the SM Higgs, so the efficiency drops rapidly as we depart from the SM Higgs mass (as shown in the right panel of Fig.~\ref{bbgaga_eff}).
Therefore, we use the following invariant mass cuts:
\begin{eqnarray}
|m_{\gamma\gamma} - m_{\phi}| < 3\mbox{ GeV},\ \ \ |m_{b\bar b} - m_{\phi}| < 25\mbox{ GeV},
\end{eqnarray}
which reduce to the cuts from Ref.~\cite{ATL-PHYS-PUB-2017-001} for $m_\phi=125\mbox{ GeV}$.
The right panel in Fig.~\ref{bbgaga_eff} shows the total signal efficiency as a function of $m_\phi$, for $m_\phi=125\mbox{ GeV}$ (\emph{SM di-Higgs cuts}) and for variable $m_\phi$ (\emph{modified cuts}).
The efficiencies obtained from our Monte Carlo simulation have been scaled to match the efficiency obtained in Ref.~\cite{ATL-PHYS-PUB-2017-001} for the point $m_\phi=125\mbox{ GeV}$ ($\approx 2.89\%$).

Using the signal efficiency with modified cuts, and the branching fractions shown in Fig.~\ref{phibr_photon}, we find 95\% CL limits on $g_{h\phi\phi}$.
To find these limits, we scale the backgrounds from the SM di-Higgs search from Ref.~\cite{ATL-PHYS-PUB-2017-001}.
This is a conservative estimate because the continuum diphoton background is expected to be much smaller for higher invariant mass regions.
We find that current di-Higgs searches in the $b\bar b \gamma\gamma$ final state can be straightforwardly adapted to be sensitive to the exotic triple Higgs coupling $g_{h\phi\phi}$ for $m_\phi \lesssim 250\mbox{ GeV}$ ($300\ifb$) and $m_\phi \lesssim 300\mbox{ GeV}$ ($3000\ifb$).

\section{Constraining $hAA$ coupling}

The lightest almost inert neutral state can also be the pseudoscalar $A$.
How $A$ decays largely depends on the relative size between the $\mathbb{Z}_2$-breaking parameters $\ep_A$ and $\ep_V$. For instance,
as shown in Fig.~\ref{Abr}, for $\ep_V \approx \epsilon_A$, the decay of $A$ is dominated by $b\bar{b}$ for $m_A<125$ GeV and by $Zh$ for $m_A>125$ GeV.
Therefore, the $4b$ channel previously studied (section~\ref{sec:4b}) also constrains $g_{hAA}$.
The new ingredient here is that for high masses of $A$, the pair produced $A$s can undergo cascade decays to two $Z$ bosons and two SM $h$s.
If all of these particles are produced on shell, it is possible to reconstruct the entire decay chain. 

Each subsection below assumes a distinct relative size between $\ep_V$ and $\epsilon_A$, and discusses possible constraints on $g_{hAA}$ arising from the dominant decays.

\subsection{$\ep_A\gtrsim\ep_V: AA\to (b\bar b)(b\bar b)$}

As mentioned before, for a wide region of the parameter space, $A\to b\bar b$ is one of the main decay modes of $A$.
In general, this is the dominant decay mode (almost $100$\%) for $\epsilon_A \gtrsim \epsilon_V$.
Also, for $\epsilon_A\approx \epsilon_V$, this mode dominates for $m_A\lesssim 215\mbox{ GeV}$, as shown in Fig.~\ref{Abr}.
Therefore, the limits obtained in Fig.~\ref{limit4b} for $g_{h\phi\phi}$ in the $(b\bar b)(b\bar b)$ final state are also valid for $g_{hAA}$.

\subsection{$\ep_A\lesssim\ep_V: AA \to (Zh)(Zh)$}
The final states coming from a pair of cascade decays can be rich and complex. In order to reconstruct the decay chain, we require $h\to b\bar{b}$ for both SM Higgses and at least one $Z$ boson undergoing leptonic decay. 
Depending on how the other $Z$ boson decays, there arise multiple decay channels that can be potentially interesting. 

The cleanest channel is the one with both $Z$ bosons decaying leptonically, but it suffers from a tiny signal rate.
The second cleanest channel is when $ZZ\to (\ell\bar{\ell})(\nu\bar{\nu})$, which has a same final state as the dileptonic decay of $t\bar{t}h$, where $h\to b\bar{b}$.
There exist dedicated search on $t\bar{t}h$ from CMS \cite{CMS:2018alh}, where many of the search strategies can be migrated here.
However, a crucial difference between $t\bar{t}h$ and our model is that the former requires a $Z$ veto which is undesirable in the latter case. 
In order to constrain the coupling from this channel, we estimate the SM backgrounds based on our own simulation.
We also considered $ZZ\to(\ell\bar{\ell})(jj)$, but the SM backgrounds such as $Z+$jets are difficult to reduce. 

Therefore, we focus on the final state $(\ell\bar{\ell} b\bar{b})(\nu\bar{\nu}b\bar{b})$, coming from the cascade decays, $A\rightarrow Zh$, of pair produced $A$s at the 13 TeV LHC. The dominating backgrounds include $t\bar{t}$, single top and $Z+$jets, as well as other SM rare processes (cf. Table~\ref{tab:llbbbb}).
We first perform a detailed benchmark study, for which we generate the number of background events that is equivalent to an integrated luminosity of 300 fb$^{-1}$.

\begin{table}[h!]
   \centering
     \begin{tabular}{ |l | c c c c c |} 
    \hline
 $AA\to(Zh)(Zh)$ & $\sigma$(fb) & Initial @ & $2\ell 4b$  & on-$Z$   & $M_{T2W}>100$ \\ 
$\quad\quad\to(\ell\ell bb)(\nu\nu b b)$ &  & 300 fb$^{-1}$ & & $\met >120$ & $M_{T2t}>200$ \\
 \hline
BM1: $m_A=220$ &0.2 &60 &2.5 &1.9 &1.4 \\ [1ex]
BM2: $m_A=315$ &0.08 &24 & 1.1&0.8 &0.6 \\ [1ex]
 \hline
$t\bar t + jj \to(\nu\ell b)(\nu\ell b)+jj$ & 19680 & $5.9\times10^6$ & 8330  &  182& 0.9\\  [1ex] 

$Z+bbjj$, $Z \to \ell\ell$  & 19790 &$5.9\times10^6$  & 1275&86 & 3.6\\ [1ex] 

$tW + b\bar b \to (\nu\ell b)(\nu\ell)+b\bar b$ & 145 & 43500 & 443&14  & 0\\ [1ex] 

 $t\bar tV \to(\nu\ell b)(\nu\ell b)+V$ & 43 & 12900 & 42&  2& 0\\ [1ex] 

 $tth\to(\nu\ell b)(\nu\ell b)(bb)$ & 15 & 4500 &122 & 3& 0\\ 
 \hline
 SM Backgrounds Total & - & - & - & -&  4.5\\
 \hline
\end{tabular}
    \caption{Signal cross section is given by $\sigma\big(pp\to AA\to(Zh)(Zh)\big)\times 2\mbox{Br}(Z\to\ell\bar{\ell})\mbox{Br}(Z\to\nu\bar{\nu})\times\mbox{Br}(h\to b\bar{b})^2$, where $g_{hAA}$ is taken to be $4m_A^2/v$ in the signal benchmarks. $j= u, c, d, s, b, g$.}
    \label{tab:llbbbb}
\end{table}

Table~\ref{tab:llbbbb} shows the cutflow for two signal benchmarks and the main SM backgrounds.
For the pre-selection, we require 2 leptons and 4 $b$-tagged jets with standard $p_T$ and $\eta$ cuts.
To suppress $Z+$jets, we require $\met>120$ GeV.
To suppress leptonically-decayed top backgrounds, we further require that the lepton pair has an invariant mass within the 15 GeV window around the $Z$ boson mass (on-$Z$). 
Furthermore, we investigate $M_{T2}$ type of kinematic variables \cite{hep-ph/9906349,1411.4312}.
In particular, $M_{T2W}$, formed by two leptons plus $\met$, and $M_{T2t}$, formed by two leptons and two leading $b-$tagged jets plus $\met$, can be used to discriminate the di-leptonic $t\bar{t}$.
Even though the efficiency for pre-selection is $\lesssim5\%$ for signals, the cut efficiency is $\gtrsim50\%$ in spite of different mass values.

\begin{figure}[h!]
\begin{center}
\begin{subfigure}[b]{0.48\textwidth}
\includegraphics[width=\textwidth]{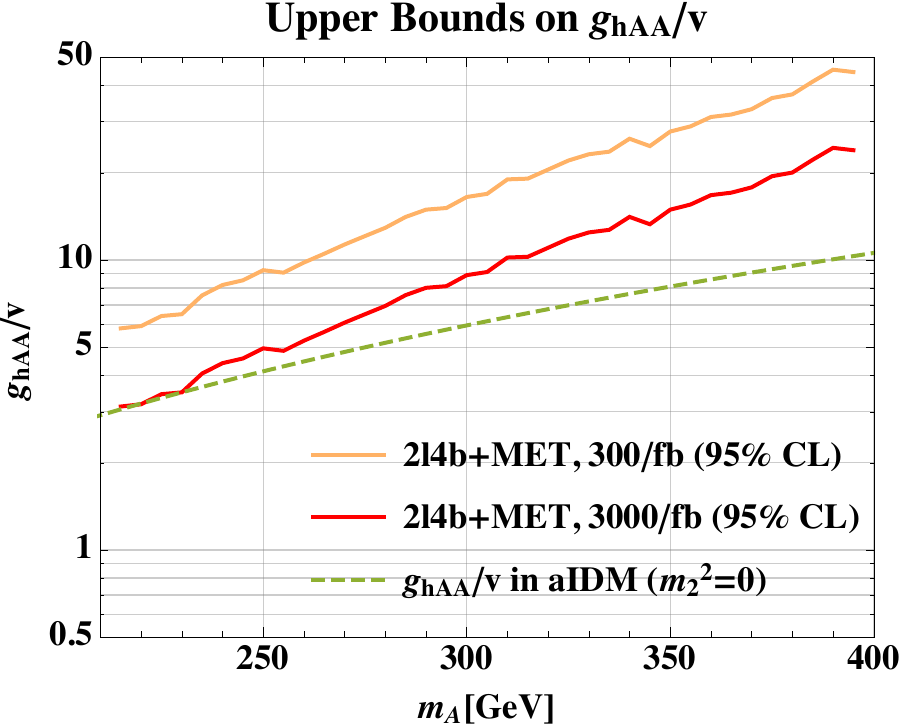}
\end{subfigure}
\hfill
\begin{subfigure}[b]{0.47\textwidth}
\includegraphics[width=\textwidth]{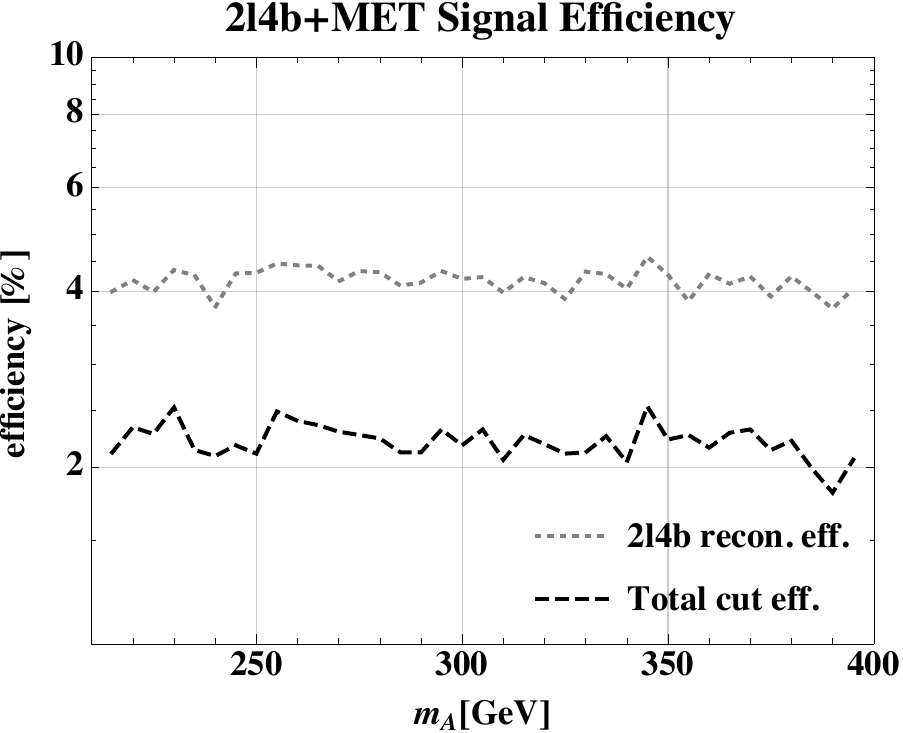}
\end{subfigure}
\caption{Left panel shows the upper limits on $g_{hAA}/v$ from $pp\rightarrow h \rightarrow AA\rightarrow (Zh)(Zh)\rightarrow(\ell\ell bb)(\nu\nu bb)$ at the LHC with $300\mbox{ fb}^{-1}$ and HL-LHC ($3000\mbox{ fb}^{-1}$).
We also include the value of $g_{hAA}/v$ in the aIDM for $m_2^2=0$, which is given by $4m_{A}^2/v^2$.
Right panel shows the cut efficiency for the signals.}\label{fig:llbbbb}
\end{center}
\end{figure}

From Fig.~\ref{fig:llbbbb}, we see that the HL-LHC will be sensitive in the region around $215-230\mbox{ GeV}$.
As well as for the other modes, this reach could be improved with a more sophisticated analysis, therefore this remains an interesting final state to look at.

\section{Results and Conclusions}

In this work, we assume a simple extended Higgs sector (2HDM) that exhibits alignment without the requirement for the additional scalars to be much heavier than the electroweak scale. 
A mild hierarchy ($m_{\phi,A,H^\pm}\gtrsim m_h$) allows the additional states to have large couplings to the Standard Model Higgs.
We investigate the potential of the LHC for constraining the exotic triple Higgs couplings $hHH$, where $h$ represents the SM Higgs and $H$ the lightest exotic neutral Higgs. 
In particular, the exotic Higgs sector is assumed to be odd under an approximate $\mathbb Z_2$ symmetry, such that single production of the additional states is highly suppressed, leaving pair production as the main production mechanism.
Motivated by the non-decoupling feature exhibited by this scenario, where the triple Higgs coupling $g_{hHH}$ scales roughly as the mass squared of $H$, we look at the pair production of $H$ via an off-shell SM Higgs: $pp\rightarrow h^* \rightarrow HH$.
This cross section does not drop as fast as the square of the gluon parton distribution function, because of the enhancement of the coupling in the high mass regime.

We identify the main decay modes of pair produced $H$s by varying the size of the small $\mathbb Z_2$-breaking parameters $\ep_V$ and $\ep_A$, that control the tree-level couplings of $H$ to SM vector bosons and fermions, respectively.
Furthermore, when both $\ep_V$ and $\ep_A$ become vanishingly small, the loop-induced di-photon decay becomes dominant for the lightest exotic neutral Higgs. 
\begin{figure}[h!]
\begin{center}
\begin{subfigure}[b]{0.48\textwidth}
\includegraphics[width=\textwidth]{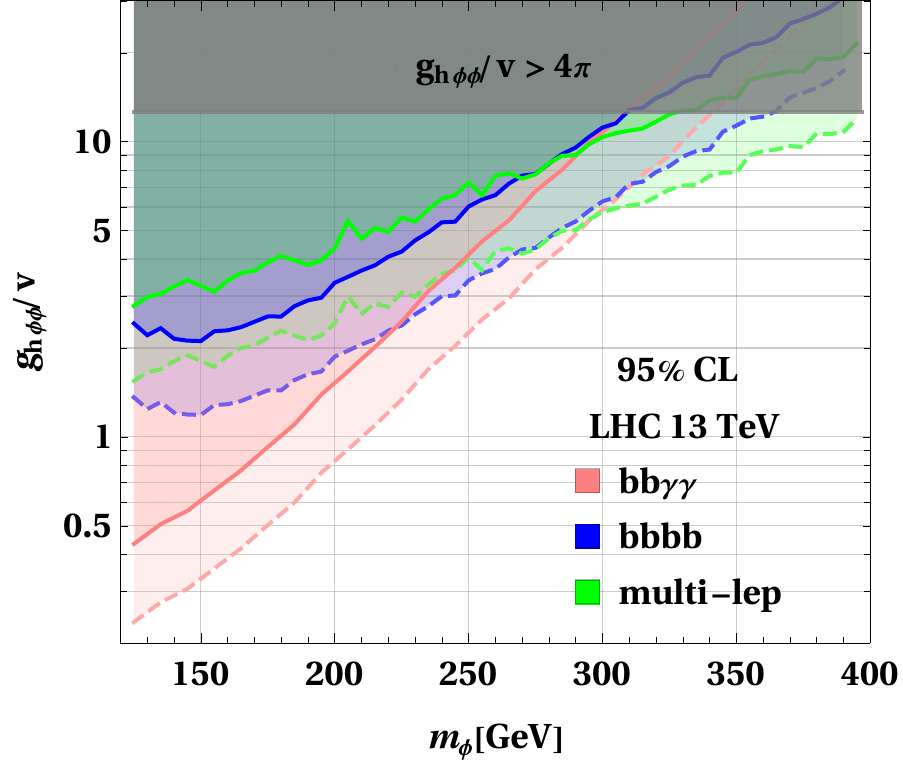}
\end{subfigure}
\hfill
\begin{subfigure}[b]{0.48\textwidth}
\includegraphics[width=\textwidth]{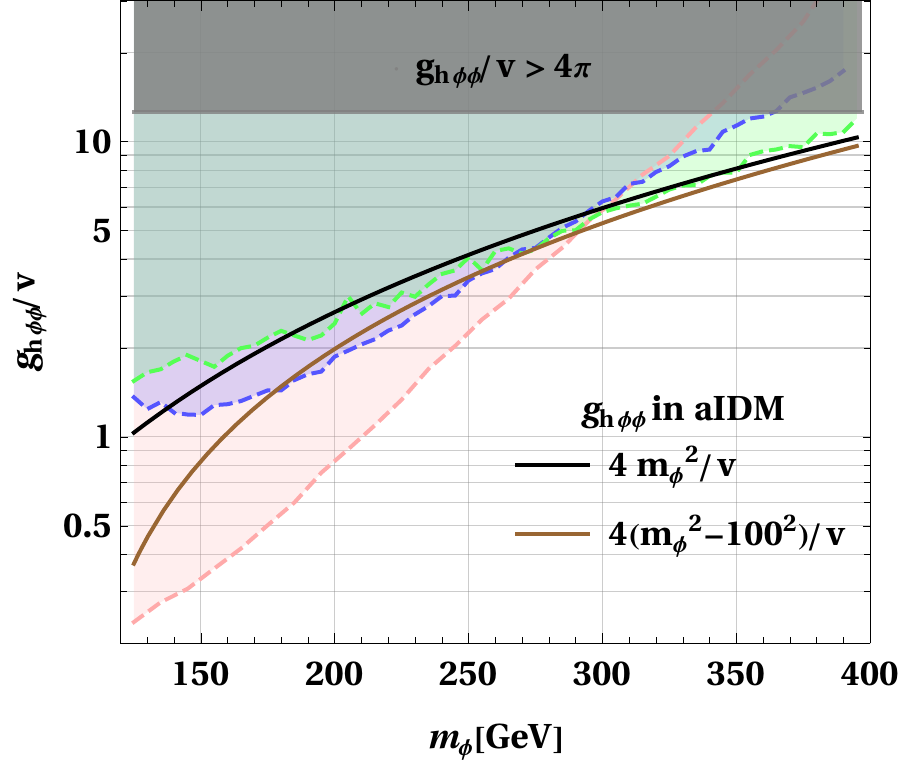}
\end{subfigure}
\caption{Combined 95\% CL upper bounds on $g_{h\phi\phi}$ at 13 TeV LHC.
The solid (dashed) curve represents the exclusion limit at $300\mbox{ fb}^{-1}$ ($3000\mbox{ fb}^{-1}$).
Examples of $g_{h\phi\phi}$ values expected from the almost inert doublet model are plotted in the right panel.}\label{fig:summary}
\end{center}
\end{figure}

We find that the HL-LHC can be efficient in constraining $g_{hHH}$ in a non-decoupling extended Higgs sector.
In particular, the most sensitive channel is $4b$-jets when $\ep_A>\ep_V$, and $3\ell$ when $\ep_A<\ep_V$, both of which are covered by existing ATLAS and CMS searches.
Fig.~\ref{fig:summary} shows a summary of the regions, in the $g_{h\phi\phi}$ vs. $m_\phi$ plane, excluded by the different final states we analyzed.
For $b\bar b b\bar b$ and $b\bar b\gamma\gamma$, we extrapolated the backgrounds from the SM di-Higgs searches, that are not optimized for the heavy Higgses. Consequently, the limit on the coupling obtained in these analyses is rather conservative, and has the potential to be considerably improved by a full optimized study.

In this work we have analyzed the scenario where the exotic Higgses decay promptly inside the detector.
However, this is no longer valid if all the $\mathbb Z_2$-breaking parameters become $\lesssim\mathcal{O}(10^{-4})$, where searches based on displaced jets, leptons, or delayed photons become important.
We leave this non-prompt scenario for future work, since it features a very distinctive phenomenology and requires different search strategies.

\newpage

\appendix{Appendix A: Details of Benchmark Studies}

\subsection{4 $b$ channel}\label{app:bbbb}

Following the analysis in Ref.~\cite{Aaboud:2018knk}, we require four $b$-tagged jets with $p_T>40\mbox{ GeV}$.
The pairings of $b$-jets with Higgs boson candidates are accepted only if they satisfy
\begin{eqnarray}
\begin{drcases}
\frac{360~\GeV}{m_{4j}} - 0.5 < \Delta R_{jj, \mathrm{lead}} < \frac{653~\GeV}{m_{4j}} + 0.475\\
\frac{235~\GeV}{m_{4j}}\qquad\ < \Delta R_{jj, \mathrm{subl}} < \frac{875~\GeV}{m_{4j}} + 0.35
\end{drcases}
\mathrm{if}\ m_{4j} < 1250~\GeV,
\end{eqnarray}
\begin{eqnarray}
\begin{drcases}
0< \Delta R_{jj, \mathrm{lead}} < 1\\
0 < \Delta R_{jj, \mathrm{subl}} < 1
\end{drcases}
\mathrm{if}\ m_{4j} > 1250~\GeV.
\end{eqnarray}
If more than one $b$-jet pairing satisfy these conditions, the one with the smallest value $D_{HH}$ value is chosen, where
\begin{eqnarray}
D_{HH} = \frac{\left|m_{2j}^\mathrm{lead} - \frac{120}{110} m_{2j}^\mathrm{subl} \right|}{\sqrt{1+\left(\frac{120}{110}\right)^{2}}}.
\end{eqnarray}
Additionally, the $p_T$ of the leading and subleading Higgs candidate is required to have
\[
p_T^\mathrm{lead} > 0.5 m_{4j} - 103~\GeV,
\]
\[
p_T^\mathrm{subl} > 0.33 m_{4j} - 73~\GeV.
\]
The pseudorapidity difference between the Higgs boson candidates is required to satisfy $|\Delta \eta_{HH}|<1.5$.
Finally, the Higgs boson candidates are required to satisfy
\begin{equation}
X_{HH} = \sqrt{\left(\frac{ m_{2j}^\mathrm{lead} - 120~\GeV}{0.1m_{2j}^\mathrm{lead}}\right)^2 + \left(\frac{m_{2j}^\mathrm{subl} - 110~\GeV}{0.1m_{2j}^\mathrm{subl}}\right)^2} < 1.6.
\end{equation}

\subsection{multi-$\ell$ channel}\label{app:multil}

For triggering, we require at least 2 jets with transverse momentum ($p_T$) larger than 30 GeV, falling within $|\eta|<2.4$, and missing transverse energy (MET) greater than 50 GeV. We veto events with $b-$tagged jets.

At least 3 isolated electrons or muons are required to have 
\begin{eqnarray}
\begin{drcases}
\quad\,\,\,\mathrm{leading} \, e\, (\mu): p_T>25\,(20)~\GeV\\
\mathrm{subleading}\, e\, (\mu): p_T>15\,(10)~\GeV
\end{drcases}
\mathrm{if}\ \mathrm{H_T} < 400~\GeV,
\end{eqnarray}
\begin{eqnarray}
\begin{drcases}
\quad\,\,\,\mathrm{leading} \, e\, (\mu): p_T>15\,(10)~\GeV\\
\mathrm{subleading}\, e\, (\mu): p_T>15\,(10)~\GeV
\end{drcases}
\mathrm{if}\ \mathrm{H_T} > 400~\GeV.
\end{eqnarray}
The lowest-$p_T$ lepton must have $p_T>10$ GeV in both cases.
We further veto events containing an opposite-charge, same-flavor lepton pair with an invariant mass within the 15 GeV window around the $Z$ boson mass.

The search is then categorized based on different $\met$, $H_T$ and $M_T$ bins, following \cite{Sirunyan:2017hvp}, where $M_T$ is defined to be the minimum of
\[
M_T =\sqrt{2p_\text{T}^{\ell}p_\text{T}^{\text{miss}}[1-\cos(\phi_{\ell}-\phi_{\vec{p}_\text{T}^{\text{miss}}})]}
\]
out of all combinations.
The bin that is most sensitive to our signal is the one with 150 GeV$<\met<300$ GeV, $H_T>400$ GeV, and $M_T<120$ GeV.

\subsection{$b\bar{b}\gamma\gamma$ channel}

Following the analysis in Ref.~\cite{ATL-PHYS-PUB-2017-001}, we require 2 photons and between 2 and 5 jets with $|\eta|<2.5$ and $\pt>30\mbox{ GeV}$. The leading b-jet satisfies $\pt>40\mbox{ GeV}$, while for photons $|\eta|<2.37$ and $\pt>30\mbox{ GeV}$.
Events containing isolated leptons with $\pt > 25\mbox{ GeV}$ are vetoed.
The angular separation $\Delta R = \sqrt{\Delta\eta+\Delta\phi}$ satisfies
\begin{eqnarray}
    0.4 < \Delta R_{b\bar b} < 2.0,\nonumber\\
    0.4 < \Delta R_{\ga \ga} < 2.0,\\
    0.4 < \Delta R_{\ga, jet}.\nonumber
\end{eqnarray}
The two photon and two b-jet systems fulfil $p_T^{\ga\ga},p_T^{bb}>80\mbox{ GeV}$.

\section*{Acknowledgements}
We thank M. A.~Luty for the inspiration of this work. It is CG's pleasure to thank Y.~Bai, B.~Dobrescu, P.~Fox, D.~Liu, M.~Low, P.~Machado for useful discussions and comments on the manuscript.
NN was supported by FONDECYT (Chile) grant 3170906 and in part by Conicyt PIA/Basal FB0821. CG was supported by
Fermilab, operated by Fermi Research Alliance,LLC under contract number
DE-AC02-07CH11359 with the United States Department of Energy.

\end{document}